# Probing the spin-polarized electronic band structure in monolayer transition metal dichalcogenides by optical spectroscopy

Zefang Wang[1], Liang Zhao[2], Kin Fai Mak[1*], and Jie Shan[1,2*]

[1] Department of Physics and Center for 2-Dimensional and Layered Materials, The Pennsylvania State University, University Park, Pennsylvania 16802-6300, USA

[2] Department of Physics, Case Western Reserve University, Cleveland, Ohio 44106-7079, USA

[*] Fax: 1(814) 865-0978; Email: kzm11@psu.edu (K.F.M.); jus59@psu.edu (J.S.)

**Abstract**

We study the electronic band structure in the K/K' valleys of the Brillouin zone of monolayer $WSe_2$ and $MoSe_2$ by optical reflection and photoluminescence spectroscopy on dual-gated field-effect devices. Our experiment reveals the distinct spin polarization in the conduction bands of these compounds by a systematic study of the doping dependence of the A and B excitonic resonances. Electrons in the highest-energy valence band and the lowest-energy conduction band have antiparallel spins in monolayer $WSe_2$, and parallel spins in monolayer $MoSe_2$. The spin splitting is determined to be hundreds of meV for the valence bands and tens of meV for the conduction bands, which are in good agreement with first principles calculations. These values also suggest that both *n*- and *p*-type $WSe_2$ and $MoSe_2$ can be relevant for spin- and valley-based applications.

Two-dimensional (2D) transition metal dichalcogenides (TMDs) have attracted significant interest due to their unique electronic and optical properties that arise from the emergent valley degree of freedom and the strong spin-orbit interactions (SOIs) [1-3]. These properties have also been actively explored for new electronics and optoelectronics applications [4-15]. In the monolayer limit, group-VI TMDs (MX$_2$, M = Mo, W; X = S, Se) are direct band-gap semiconductors with the fundamental energy gap located at the K and K' points of the Brillouin zone [16, 17]. Because of the broken inversion symmetry and out-of-plane mirror symmetry, the strong SOIs split the bands in the K/K' valley into the out-of-plane spin-up and spin-down bands in monolayer TMDs [1, 2]. The spins of the degenerate bands at the two valleys are opposite, required by time reversal symmetry. This is known as spin-valley or spin-momentum locking [1], which plays a crucial role in the long valley and spin lifetimes in monolayer TMDs [18, 19]. The spin splitting of the highest-energy valence bands at the K/K' points ($\Delta_v$) is known to be large (100's meV) [1]. In contrast, spin splitting of the lowest-energy conduction bands ($\Delta_c$) is expected to be much smaller since the bands consist mostly of the $d_{z^2}$-orbitals of the metal atom, for which SOIs vanish [1]. Understanding the spin-polarized conduction band structure is important for a wide array of spin-related phenomena in monolayer TMDs. Examples include the exciton optical selection rules [1, 3, 20, 21], exciton and spin lifetimes [18, 21-23], and the spin and valley Hall effects [1, 12, 15]. A large $\Delta_c$ is also important for spintronics applications based on $n$-type TMDs. Recent first principles calculations including higher-order effects have predicted a finite $\Delta_c$ with opposite sign for Mo- and W-based TMDs [24-26] [i.e. order of spin-up and spin-down bands, Fig. 1(a), 1(b)]. Meanwhile, studies on the sign of the exciton g-factor [27-31] and the dark exciton state [21, 32-34] have shown evidence supporting the theory. A direct experimental probe of the spin-polarized band structure and measurement of the magnitude of $\Delta_c$ in monolayer TMDs are, however, still lacking.

In this Letter, we directly probe the spin-polarized band structure of monolayer WSe$_2$ and MoSe$_2$ by optical reflectance and photoluminescence (PL) spectroscopy on high-quality dual-gated field-effect devices. The method relies on the optical selection rules that interband transitions in the dipole approximation are allowed only between bands of the same electron spin and valley [1, 3, 20]. We measure the energy shift of the absorption resonances and the PL Stokes shift while the Fermi level is continuously scanned from the upper valence bands to the upper conduction bands by dual local gates. Distinct doping dependences as a result of Pauli blocking and renormalization of the band gap and exciton binding energies have been observed for monolayer WSe$_2$ and MoSe$_2$. These observations form a comprehensive picture of the distinct spin-polarized band structures of W- and Mo-based compounds. Based on this picture, we estimate the conduction and valence band spin splitting from the Fermi level shift for WSe$_2$ and MoSe$_2$ to be 10's meV and of opposite sign, which are in good agreement with the results of ab initio calculations [24, 25]. We note, however, that the electronic transitions probed by optical absorption and PL are modified by the Coulomb interaction effects, which are not considered in the simple density functional theory (DFT) calculations [24, 25]. A more accurate description of our results and determination of the band spin splitting requires a many-body theory that takes into account the interaction effects [26, 35, 36].

Figure 1(d) illustrates the schematic structure of dual-gate field-effect devices of monolayer TMDs employed in this study. It consists of a TMD monolayer encapsulated by hexagonal boron nitride (hBN) thin films (~ 15 – 20 nm on both sides), and few-layer graphene as both gate and contact electrodes. The multilayer devices were fabricated based on the layer-by-layer dry transfer method developed by Wang et al. [37]. In short, thin flakes of TMDs, hBN and graphene were first mechanically exfoliated from their bulk crystals onto silicon substrates covered by a 300-nm thermal oxide layer. The thickness of the thin flakes was first estimated by their optical reflectance contrast and then confirmed by PL spectroscopy for TMDs or atomic force microscopy (AFM) for others. They were then picked layer-by-layer by a polypropylene carbonate (PPC) film on a polydimethyl siloxane (PDMS) stamp. The completed multilayer stacks were then released onto silicon substrates with pre-patterned gold electrodes. The stack was aligned such that each graphene electrode makes contact with only one gold electrode. (See Supplementary Information for device images.) The PPC residual was dissolved in anisole before measurements. Because of the full encapsulation by hBN, the TMD samples are typically of high quality as shown below.

The optical measurements were performed at 3 K in an Attocube cryostat. For reflection spectroscopy, broadband radiation from a super-continuum laser was focused onto the samples with a spot size of ~ 1 μm by a microscope objective. The reflected light was collected by the same objective and dispersed by a grating spectrometer equipped with a charge-coupled device (CCD). The reflectance contrast at photon energy $E$ was determined as $\frac{\Delta R}{R} \equiv \frac{I-I_0}{I_0}$, where $I(E)$ and $I_0(E)$ are reflectance spectrum acquired from the multilayer stack with and without the TMDs, respectively. For PL spectroscopy, a He-Ne laser at 632.8 nm was used as the excitation source and a long-pass filter was introduced in front of the spectrometer to block the laser radiation for PL detection. In all measurements, the illumination power on the samples was kept below 100 μW to limit sample heating.

The dual-gate structure allows us to independently control the doping density and the vertical electric field on monolayer TMDs. Figure 2(a) and 2(b) illustrate several representative PL spectra of monolayer WSe$_2$ under a combination of the top and back gate voltages. Aside from narrow emission lines arisen from localized excitons at low energies, the PL spectrum is dominated by emission of the neutral exciton (A$^0$) and charged excitons (A'). The emergence of the positively or negatively charged excitons and reduction of the neutral exciton PL intensity are a result of *p*- or *n*-doping in WSe$_2$, as reported in earlier studies [8, 38, 39]. The charged exciton emission has a peak width of ~ 5 meV, illustrating the high sample quality. We use the maximum PL intensity of the neutral exciton as a criterion to determine the gate voltages for zero doping. Figure 2(c) shows the top gate voltage $V_T$ required to keep doping at zero for a given back gate voltage $V_B$ (symbols). As expected, this combination of gate voltages follows a straight line (orange). The slope of the line depends on the relative thickness of the top and bottom hBN layers. Along this line the electric field on the sample varies while the doping level remains close to zero. Figure 2(d) shows several PL spectra near this zero-doping line. The electric-field effects on the neutral exciton are small. Next, we construct the zero-field line from the zero-doping line as its mirror image with respect to the $V_T$ axis and passed through the origin [blue line in Fig. 2(c)]. Along this line, doping in the sample varies (with the upper right and lower left part corresponding to *n*- and *p*-doping, respectively), while the vertical electric field is fixed approximately at zero. The measurements below were restricted along this line.

We calibrated the doping density $n = n_i + C_T V_T + C_B V_B$ using the applied gate voltages and the top and back gate capacitances: $C_T = \frac{\varepsilon \varepsilon_0}{d_T}$ and $C_B = \frac{\varepsilon \varepsilon_0}{d_B}$ ($\varepsilon_0$ is the vacuum permittivity). The latter were determined from the hBN dielectric constant ($\varepsilon = 2.5$ [40]) and thickness ($d_T$ and $d_B$) from the AFM measurements. In our devices the top and back gates are nearly identical and the doping density is thus determined by the sum of the gate voltages $V_G = V_T + V_B$. Constant $n_i$ was set to $n = 0$ at the gate voltages for which the Fermi level touches the conduction or the valence band as discussed below. Positive $n$'s are for the electron density and negative $n$'s for the hole density.

The raw reflectance contrast spectra $\frac{\Delta R}{R}$ of monolayer WSe$_2$ at different gate voltages $V_G$ are shown in the Supplementary Information. Since the real and imaginary parts of the optical conductivity of the sample are mixed due to optical interference in the multilayer device on Si/SiO$_2$, the reflectance contrast spectrum is not equivalent to the sample's absorption spectrum and will only be used to track the shift with doping of the resonances. The uncertainty associated with the absolute peak energy determined from this method was estimated to be less than the transition linewidth, i.e. < 5-15 meV (depending on $n$). To further suppress the background, we computed the photon energy derivative of the reflectance contrast spectrum [Fig. 3(a) and 3(b)]. The vertical-line features (doping independent) are artifacts from normalization of the signal and reference reflectance spectra taken at two different sample locations. The main spectral features are the A and B excitons. They correspond to the optical transitions from the two spin-split valence bands to the corresponding conduction bands [1, 3, 20] [Fig. 1(a)], modified by the strong *e-h* interactions [41-45]. Near $V_G = 0$, the spectra in Fig. 3(a) are dominated by the neutral A$^0$ exciton absorption at ~ 1.73 eV (bipolar shape due to interference and derivative). With either *n*- or *p*-doping, charged exciton A' features emerge below the neutral A$^0$ exciton energy and the reflectance contrast of the A$^0$ exciton vanishes. We note that the high sample quality has allowed us to observe the splitting of the previously reported negative trion [46] for $0 < n \leq \sim 10^{12}$ cm$^{-2}$. Our observation is consistent with the prediction of exchange interaction-induced splitting of trions [46], but further investigations are warranted to fully understand their origin. Similarly, the neutral B$^0$ exciton also vanishes while the charged B' excitons emerge [Fig. 3(b)]. It has been shown previously that at low temperatures, charged excitons are formed when the Fermi level crosses the conduction band minimum or the valence band maximum [8, 38, 39]. We set $n = 0$ at the two gate voltages, where the reflectance contrast of neutral A$^0$ exciton just vanishes (dashed lines). The gate-voltage interval between these two lines corresponds to moving the Fermi level across the band gap of the material. With further electron doping, the A' feature redshifts, and the B' feature blueshifts. On the other hand, with hole doping, the A' feature blueshifts and the B' feature redshifts. These doping dependences of the A' and B' peak energies ($E_{A'}$, $E_{B'}$) are summarized in Fig. 4(a).

To understand the above findings, we consider the mechanisms that can influence the optical spectrum including elastic Coulomb scattering, screening, and Pauli blocking. While elastic scattering of excitons with free carriers causes a spectral broadening of the excitonic features, screening and Pauli blocking cause a spectral shift [41, 47-50]. Screening of the *e-h* interactions lowers the exciton binding energy, whereas screening of the electron-electron (*e-e*) interactions induces a decrease of the quasi-particle self-energy and a band-gap renormalization to lower energies [48-50]. The net effect of screening of the Coulomb interactions for the charged exciton features is a redshift [8, 38, 39]. On the other hand, Pauli blocking due to the occupation of the electronic states and the fermionic nature of the electrons leads to a blueshift of the resonances. Our findings [Fig. 4(a)] can thus be explained based on the spin-polarized band structure of Fig. 1(a), where the lower conduction band and the upper valence band have opposite spins. For the Fermi level between the two valence bands ($n < 0$), the B' feature is affected only by the screening effect and redshifts; the A' feature is affected by both screening and Pauli blocking with the latter dominating, and shows a net blueshift. On the other hand, for the Fermi level between the two conduction bands ($n > 0$), the behavior of the A' and B' feature is reversed, i.e. redshift for A' and blueshift for B'. For the Fermi level into the upper conduction band with further electron doping, both A' and B' are expected to be influenced by Pauli blocking and blueshift. Indeed, a careful examination of the A' feature in Fig. 3(a) and 4(a) identifies a kink at $n_0 \approx 7 \times 10^{12}$ cm$^{-2}$ followed by a very weak blueshift. Similar effect is expected for the Fermi level into the lower valence band, which requires a much higher hole doping density and is not achieved in our experiment. Although a quantitative explanation of these results requires a many-body model [48-50], which is beyond the scope of this study, we estimate the spin-splitting energy $\Delta_c$ of the conduction bands using a simple 2D ideal gas model with valley degeneracy 2 and spin degeneracy 1: $\Delta_c = \frac{\hbar^2 \pi n_0}{m_c}$. Here $m_c$ is the band mass of the lower conduction band. We obtain $\Delta_c \approx 40$ meV using $m_c \approx 0.4 m_0$ with $m_0$ denoting the free electron mass [1, 24, 25]. Combining with $E_{B'} - E_{A'} \approx \Delta_v - \Delta_c \approx 440$ meV at zero doping from experiment, we estimate $\Delta_v \approx 480$ meV. These values are in good agreement with ab initio calculations [24-26]. A similar value for $\Delta_c$ (30 ± 5 meV) was also reported from a temperature dependence study of the PL [21]. We note that the above estimates have involved several approximations including 1) ignoring the corrections from Coulomb interactions in relating $E_{B'} - E_{A'} \approx \Delta_v - \Delta_c$ [26]; 2) the use of the band mass from DFT calculations [24, 25]; 3) the uncertainty in the determination of $n_0$; and 4) ignoring the doping effects on the spin-splitting energies and the band masses.

As a comparison, we performed identical measurements on monolayer MoSe$_2$, which has a similar electronic structure but with weaker SOIs. The reflectance contrast spectra (derivatives) are shown in Fig. 3(c) and 3(d). Raw data are included in the Supplementary Information. Similar to WSe$_2$, both neutral (A$^0$ and B$^0$) and charged (A' and B') excitons can be identified, with the latter emerged with doping. The doping-induced shift of these features [summarized in Fig. 4(b)] is, however, distinct from that of WSe$_2$. Namely, with increasing both electron and hole doping, the A' feature blueshifts while the B' feature redshifts. These behaviors are compatible with the electronic band structure of Fig. 1(b), where the lower conduction band and the upper valence band have the same spins. Therefore, for the Fermi level between the conduction bands ($n > 0$) or the valence bands ($n < 0$), the A' feature is influenced by both Pauli blocking and screening, leading to a net blueshift; and the B' feature is influenced only by the screening effect and redshifts. With further electron doping when the Fermi level reaches the bottom of the upper conduction band at density $n_0 \approx 8.5 \times 10^{12}$ cm$^{-2}$ [shaded region, Fig. 3(d)], the B' feature goes through a weak kink. Similarly, we can estimate the spin-splitting energies of the bands, particularly, $E_{B'} - E_{A'} \approx \Delta_v - \Delta_c$ (where $\Delta_c$ is negative) at zero doping and $\frac{\hbar^2 \pi n_0}{m_c} = |\Delta_c|$. We obtain $\Delta_c \approx -30$ meV and $\Delta_v \approx 170$ meV (assuming $m_c \approx 0.6 m_0$ [24, 25]). These values are also consistent with ab initio calculations [24-26]. We note that the two spin-split conduction bands in monolayer MoSe$_2$ are predicted to cross at large momentums from the K/K' point due to their spin-dependent masses [24, 25]. The doping level achieved in our experiment is, however, too low to access this regime.

Finally we comment on the doping dependence of the PL process of monolayer WSe$_2$ and MoSe$_2$. In contrast to the optical absorption process, which involves interband transitions from occupied to unoccupied states and can be affected by the Pauli blocking effect upon doping, the PL (or the zero-momentum PL in the case of trions) involves recombination of e-h pairs that are relaxed to the band edges [51] [Fig. 1(c)]. The Stokes shift thus increases monotonically with doping density. The result remains valid even when the excitonic interactions are considered [52]. We included in Fig. 4(a) and 4(b) the doping dependence of the exciton PL peak energy for monolayer WSe$_2$ and MoSe$_2$, respectively. (The raw PL spectra are included in the Supplementary Information.) As expected, for MoSe$_2$ the Stokes shift increases monotonically for both n- and p-doping [51, 52]. Similar behavior is seen for WSe$_2$ for p-doping and n-doping with $n > n_0$. Negligible Stokes shift is observed for $0 < n < n_0$ since the A' exciton absorption is not affected by the Pauli blocking effect. These results further confirm the spin-polarized band structure of Fig. 1(a) and 1(b).

In conclusion, our experiment has directly probed the spin-polarized electronic band structure of group-VI TMDs and illustrated the important differences between the Mo- and W-based monolayers. The physical origin of the distinct spin polarization in the conduction bands of these materials has been revealed by ab initio calculations [24-26]. As mentioned earlier, the conduction bands mainly arise from the $d_{z^2}$-orbitals of the metal atoms, for which no spin splitting of the bands is expected. However, small $\Delta_c$ emerges when contributions from the chacogen orbitals and from other orbitals of the metal atoms are taken into account. The sign of $\Delta_c$ varies due to the competition between these two contributions. The large values of the spin-splitting energies estimated in this work suggest that both n- and p-type WSe$_2$ and MoSe$_2$ can be relevant for spin- and valley-based applications. Our study has paved the way for future exploration of spin- and valley-dependent phenomena such as the valley Hall effect and spin Hall effect in n-doped Mo-based compounds [1, 12, 15] and the dark exciton states in W-based compounds as well as the implication of the resultant long exciton lifetimes [21, 23].

**Acknowledgements**

The research was supported by the US Department of Energy, Office of Basic Energy Sciences under award no. DESC0012635 for sample and device fabrication and award no. DESC0013883 for optical spectroscopy measurements. Support for data analysis was provided by the Air Force Office of Scientific Research under grant FA9550-16-1-0249 (K.F.M.) and the National Science Foundation DMR-1410407 (J.S.).

**References:**

(1) Xiao, D.; Liu, G. B.; Feng, W. X.; Xu, X. D.; Yao, W. *Phys. Rev. Lett.* **2012,** 108, 196802.
(2) Xu, X.; Yao, W.; Xiao, D.; Heinz, T. F. *Nat. Phys.* **2014,** 10, 343-350.
(3) Mak, K. F.; Shan, J. *Nat. Photonics* **2016,** 10, 216-226.
(4) Cao, T.; Wang, G.; Han, W.; Ye, H.; Zhu, C.; Shi, J.; Niu, Q.; Tan, P.; Wang, E.; Liu, B.; Feng, J. *Nat. Commun.* **2012,** 3, 887.
(5) Zeng, H.; Dai, J.; Yao, W.; Xiao, D.; Cui, X. *Nat. Nanotechnol.* **2012,** 7, 490-493.
(6) Mak, K. F.; He, K.; Shan, J.; Heinz, T. F. *Nat. Nanotechnol.* **2012,** 7, 494-498.
(7) Sallen, G.; Bouet, L.; Marie, X.; Wang, G.; Zhu, C. R.; Han, W. P.; Lu, Y.; Tan, P. H.; Amand, T.; Liu, B. L.; Urbaszek, B. *Phys. Rev. B* **2012,** 86, 081301.
(8) Jones, A. M.; Yu, H.; Ghimire, N. J.; Wu, S.; Aivazian, G.; Ross, J. S.; Zhao, B.; Yan, J.; Mandrus, D. G.; Xiao, D.; Yao, W.; Xu, X. *Nat. Nanotechnol.* **2013,** 8, 634-638.
(9) Yuan, H.; Bahramy, M. S.; Morimoto, K.; Wu, S.; Nomura, K.; Yang, B.-J.; Shimotani, H.; Suzuki, R.; Toh, M.; Kloc, C.; Xu, X.; Arita, R.; Nagaosa, N.; Iwasa, Y. *Nat. Phys.* **2013,** 9, 563-569.
(10) Jones, A. M.; Yu, H.; Ross, J. S.; Klement, P.; Ghimire, N. J.; Yan, J.; Mandrus, D. G.; Yao, W.; Xu, X. *Nat. Phys.* **2014,** 10, 130-134.
(11) Yuan, H.; Wang, X.; Lian, B.; Zhang, H.; Fang, X.; Shen, B.; Xu, G.; Xu, Y.; Zhang, S.-C.; Hwang, H. Y.; Cui, Y. *Nat. Nanotechnol.* **2014,** 9, 851-857.
(12) Mak, K. F.; McGill, K. L.; Park, J.; McEuen, P. L. *Science* **2014,** 344, 1489-1492.
(13) Zhang, Y. J.; Oka, T.; Suzuki, R.; Ye, J. T.; Iwasa, Y. *Science* **2014,** 344, 725-728.
(14) Ye, Y.; Xiao, J.; Wang, H.; Ye, Z.; Zhu, H.; Zhao, M.; Wang, Y.; Zhao, J.; Yin, X.; Zhang, X. *Nat. Nanotechnol.* **2016,** 11, 598-602.
(15) Lee, J.; Mak, K. F.; Shan, J. *Nat. Nanotechnol.* **2016,** 11, 421-425.
(16) Splendiani, A.; Sun, L.; Zhang, Y.; Li, T.; Kim, J.; Chim, C.-Y.; Galli, G.; Wang, F. *Nano Lett.* **2010,** 10, 1271-1275.
(17) Mak, K. F.; Lee, C.; Hone, J.; Shan, J.; Heinz, T. F. *Phys. Rev. Lett.* **2010,** 105, 136805.
(18) Yang, L.; Sinitsyn, N. A.; Chen, W.; Yuan, J.; Zhang, J.; Lou, J.; Crooker, S. A. *Nat. Phys.* **2015,** 11, 830-834.
(19) Lagarde, D.; Bouet, L.; Marie, X.; Zhu, C. R.; Liu, B. L.; Amand, T.; Tan, P. H.; Urbaszek, B. *Phys. Rev. Lett.* **2014,** 112, 047401.
(20) Rose, F.; Goerbig, M. O.; Piechon, F. *Phys. Rev. B* **2013,** 88, 125438.

(21) Zhang, X.-X.; You, Y.; Zhao, S. Y. F.; Heinz, T. F. *Phys. Rev. Lett.* **2015,** 115, 257403.
(22) Wang, R.; Ruzicka, B. A.; Kumar, N.; Bellus, M. Z.; Chiu, H.-Y.; Zhao, H. *Phys. Rev. B* **2012,** 86, 045406.
(23) Song, Y.; Dery, H. *Phys. Rev. Lett.* **2013,** 111, 026601.
(24) Liu, G. B.; Shan, W. Y.; Yao, Y. G.; Yao, W.; Xiao, D. *Phys. Rev. B* **2013,** 88, 085433.
(25) Kormanyos, A.; Zolyomi, V.; Drummond, N. D.; Burkard, G. *Phys. Rev. X* **2014,** 4, 011034.
(26) Echeverry, J. P.; Urbaszek, B.; Amand, T.; Marie, X.; Gerber, I. C. *Phys. Rev. B* **2016,** 93, 121107.
(27) Li, Y. L.; Ludwig, J.; Low, T.; Chernikov, A.; Cui, X.; Arefe, G.; Kim, Y. D.; van der Zande, A. M.; Rigosi, A.; Hill, H. M.; Kim, S. H.; Hone, J.; Li, Z. Q.; Smirnov, D.; Heinz, T. F. *Phys. Rev. Lett.* **2014,** 113, 266804.
(28) MacNeill, D.; Heikes, C.; Mak, K. F.; Anderson, Z.; Kormanyos, A.; Zolyomi, V.; Park, J.; Ralph, D. C. *Phys. Rev. Lett.* **2015,** 114, 037401.
(29) Aivazian, G.; Gong, Z.; Jones, A. M.; Chu, R.-L.; Yan, J.; Mandrus, D. G.; Zhang, C.; Cobden, D.; Yao, W.; Xu, X. *Nat. Phys.* **2015,** 11, 148-152.
(30) Srivastava, A.; Sidler, M.; Allain, A. V.; Lembke, D. S.; Kis, A.; Imamoglu, A. *Nat. Phys.* **2015,** 11, 141-147.
(31) Wang, G.; Bouet, L.; Glazov, M. M.; Amand, T.; Ivchenko, E. L.; Palleau, E.; Marie, X.; Urbaszek, B. *2D Mater.* **2015,** 2, 034002.
(32) Withers, F.; Del Pozo-Zamudio, O.; Schwarz, S.; Dufferwiel, S.; Walker, P. M.; Godde, T.; Rooney, A. P.; Gholinia, A.; Woods, C. R.; Blake, P.; Haigh, S. J.; Watanabe, K.; Taniguchi, T.; Aleiner, I. L.; Geim, A. K.; Fal'ko, V. I.; Tartakovskii, A. I.; Novoselov, K. S. *Nano Lett.* **2015,** 15, 8223-8228.
(33) Wang, G.; Robert, C.; Suslu, A.; Chen, B.; Yang, S.; Alamdari, S.; Gerber, I. C.; Amand, T.; Marie, X.; Tongay, S.; Urbaszek, B. *Nat. Commun.* **2015,** 6, 10110.
(34) Arora, A.; Koperski, M.; Nogajewski, K.; Marcus, J.; Faugeras, C.; Potemski, M. *Nanoscale* **2015,** 7, 10421-10429.
(35) Qiu, D. Y.; Cao, T.; Louie, S. G. *Phys. Rev. Lett.* **2015,** 115, 176801.
(36) Crooker, S. A.; Barrick, T.; Hollingsworth, J. A.; Klimov, V. I. *Appl. Phys. Lett.* **2003,** 82, 2793-2795.
(37) Wang, L.; Meric, I.; Huang, P. Y.; Gao, Q.; Gao, Y.; Tran, H.; Taniguchi, T.; Watanabe, K.; Campos, L. M.; Muller, D. A.; Guo, J.; Kim, P.; Hone, J.; Shepard, K. L.; Dean, C. R. *Science* **2013,** 342, 614-617.
(38) Mak, K. F.; He, K.; Lee, C.; Lee, G. H.; Hone, J.; Heinz, T. F.; Shan, J. *Nat. Mater.* **2013,** 12, 207-211.
(39) Ross, J. S.; Wu, S.; Yu, H.; Ghimire, N. J.; Jones, A. M.; Aivazian, G.; Yan, J.; Mandrus, D. G.; Xiao, D.; Yao, W.; Xu, X. *Nat. Commun.* **2013,** 4, 1474.
(40) Hunt, B.; Sanchez-Yamagishi, J. D.; Young, A. F.; Yankowitz, M.; LeRoy, B. J.; Watanabe, K.; Taniguchi, T.; Moon, P.; Koshino, M.; Jarillo-Herrero, P.; Ashoori, R. C. *Science* **2013,** 340, 1427-1430.
(41) Qiu, D. Y.; da Jornada, F. H.; Louie, S. G. *Phys. Rev. Lett.* **2013,** 111, 216805.
(42) Chernikov, A.; Berkelbach, T. C.; Hill, H. M.; Rigosi, A.; Li, Y.; Aslan, O. B.; Reichman, D. R.; Hybertsen, M. S.; Heinz, T. F. *Phys. Rev. Lett.* **2014,** 113, 076802.

(43) He, K.; Kumar, N.; Zhao, L.; Wang, Z.; Mak, K. F.; Zhao, H.; Shan, J. *Phys. Rev. Lett.* **2014,** 113, 026803.
(44) Ye, Z.; Cao, T.; O'Brien, K.; Zhu, H.; Yin, X.; Wang, Y.; Louie, S. G.; Zhang, X. *Nature* **2014,** 513, 214.
(45) Wang, G.; Marie, X.; Gerber, I.; Amand, T.; Lagarde, D.; Bouet, L.; Vidal, M.; Balocchi, A.; Urbaszek, B. *Phys. Rev. Lett.* **2015,** 114, 097403.
(46) Yu, H.; Liu, G.-B.; Gong, P.; Xu, X.; Yao, W. *Nat. Commun.* **2014,** 5, 3876.
(47) Hartmut Haug; Koch, S. W., *Quantum Theory of the Optical and Electronic Properties of Semiconductors*. World Scientific Publishing Co. Pte. Ltd.: Singapore, 2004.
(48) Gao, S.; Liang, Y.; Spataru, C. D.; Yang, L. *Nano Lett.* **2016,** 16, 5568-5573.
(49) Scharf, B.; Wang, Z.; Van Tuan, D.; Shan, J.; Mak, K. F.; Zutic, I.; Dery, H. *arXiv:1606.07101* **2016**.
(50) Dery, H. *arXiv:1604.00068* **2016**.
(51) Livescu, G.; Miller, D. A. B.; Chemla, D. S.; Ramaswamy, M.; Chang, T. Y.; Sauer, N.; Gossard, A. C.; English, J. H. *IEEE J. Quantum Electron.* **1988,** 24, 1677-1689.
(52) Ruckenstein, A. E.; Schmittrink, S. *Phys. Rev. B* **1987,** 35, 7551-7557.

**Figures and figure captions:**

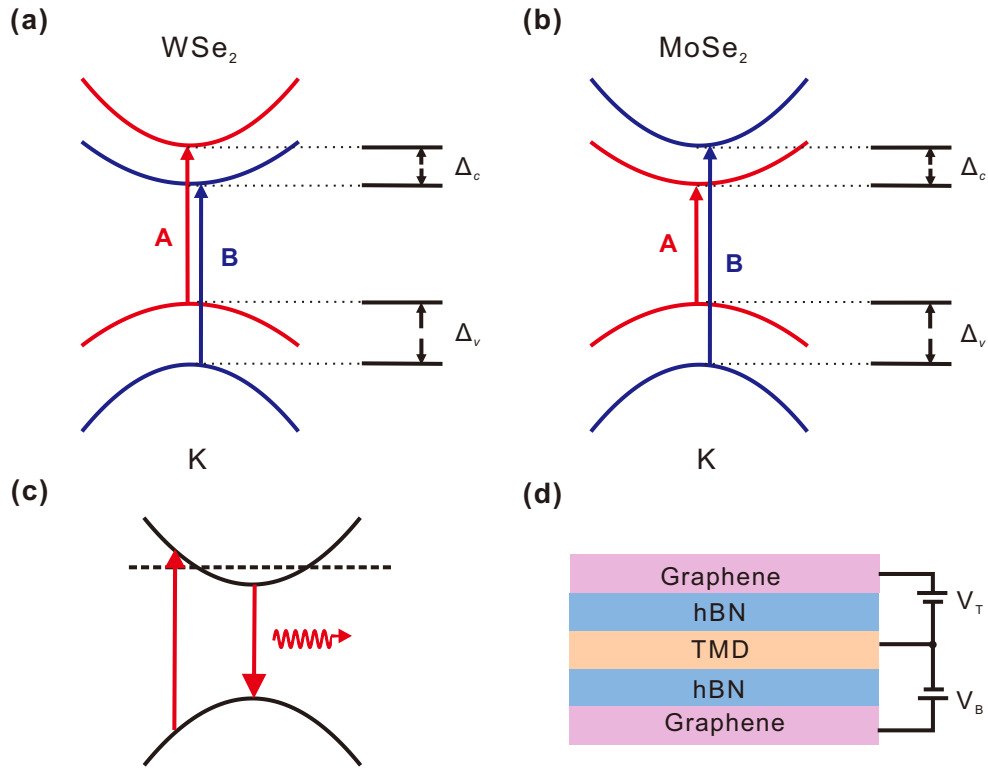

**Figure 1.** (a, b) Electronic band structure of monolayer $WSe_2$ (a) and $MoSe_2$ (b) at the K valley of the Brillouin zone. Bands of the same electron spin are shown in the same color. Arrowed lines show the optical transitions that give rise to the A (red) and B (blue) resonances. $\Delta_c$ and $\Delta_v$ are the spin-splitting energies of the lowest-energy conduction bands and the highest-energy valence bands, respectively. (c) Stokes shift of the PL energy from the absorption energy when the Fermi level (dashed line) is inside the band. (d) Schematic of a dual-gate field-effect device of monolayer TMDs with the top ($V_T$) and back ($V_B$) gate voltages applied through the graphene/hBN gates.

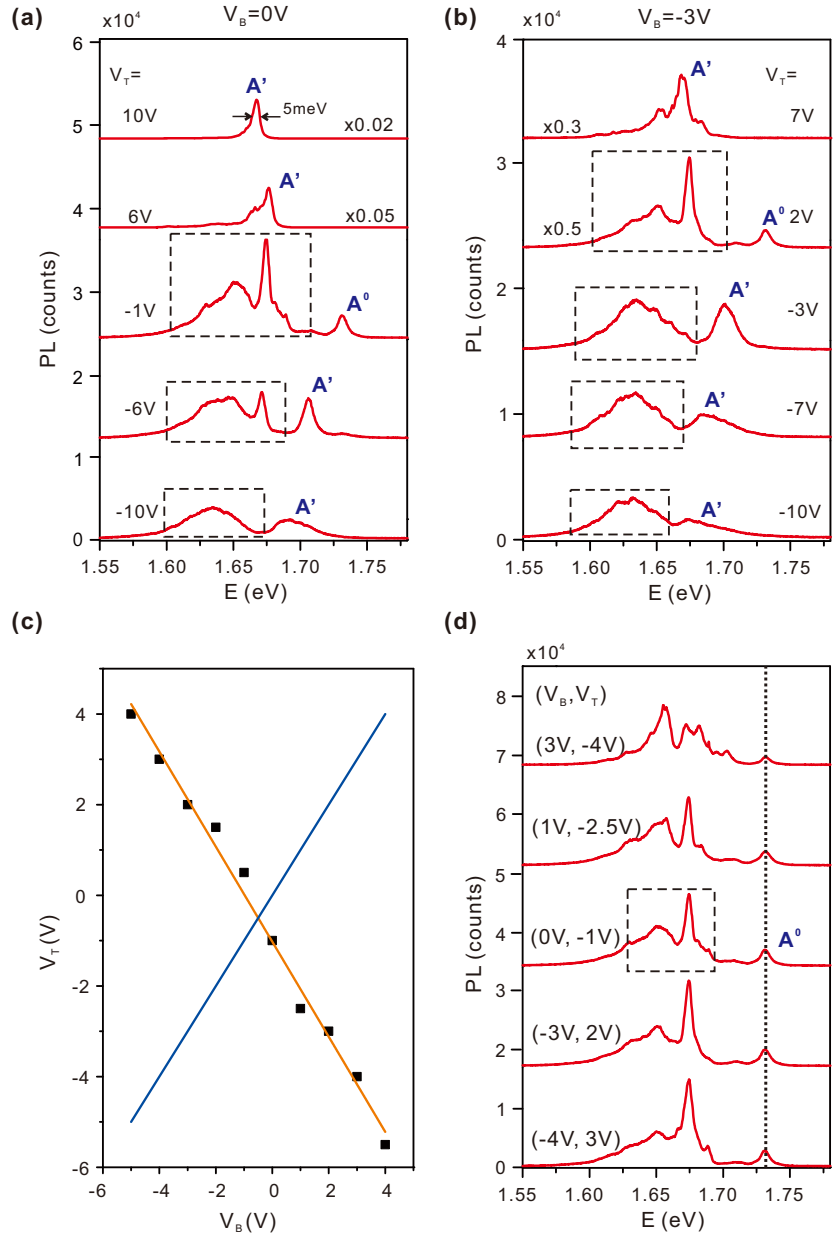

**Figure 2.** (a, b) PL spectra of monolayer WSe$_2$ under a combination of top ($V_T$) and back ($V_B$) gate voltages. Both neutral (A$^0$) and charged (A') exciton emission, as well as localized exciton emission (dashed line box) can be observed. The PL intensity of A$^0$ is quenched rapidly by gating (doping). (c) The combination of the top and back gate voltages that varies the electric field (with the doping density fixed at zero) is shown in symbols (experiment) and an orange line (linear fit). The blue line is the mirror image of the orange line with respect to the $V_T$ axis and passing through the origin. Along this line the doping density varies with the field fixed at zero. (d) Representative PL spectra measured with the gate combinations along the orange line of (c), i.e. varying the field with doping fixed approximately at zero. The dotted line marks the peak energy of A$^0$ emission and the dashed line box shows the localized exciton emission for a representative PL spectrum.

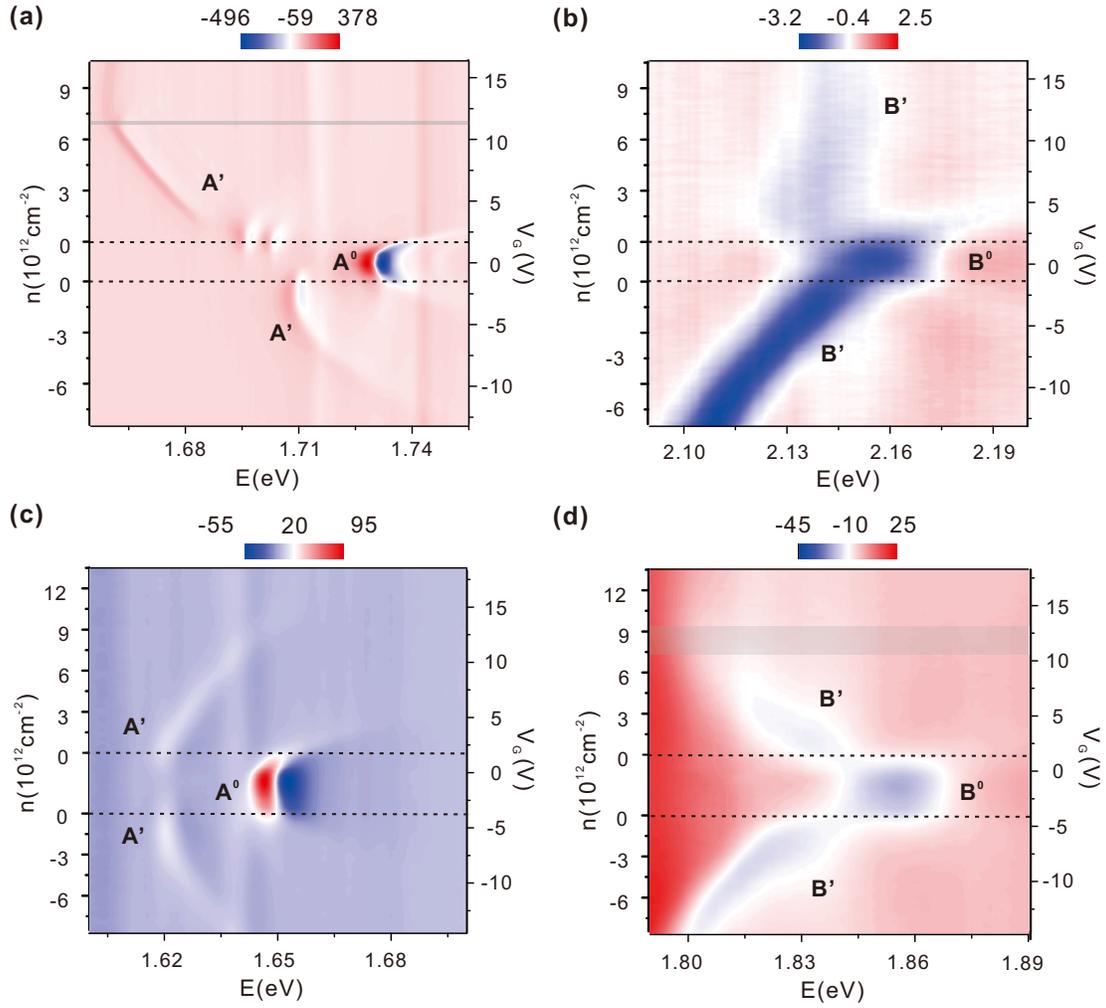

**Figure 3.** Contour plot of the derivative of the reflectance contrast spectra at varying doping densities $n$ (left axis) and $V_G$ ($= V_T + V_B$) (right axis) of WSe$_2$ (a, b) and MoSe$_2$ (c, d) centered at the A$^0$/A' resonance feature (a, c) and the B$^0$/B' resonance feature (b, d). The dashed lines correspond to zero doping. The horizontal gray shaded regions in (a) and (d) correspond to the estimated doping density $n_0$ that is required to dope into the upper conduction band.

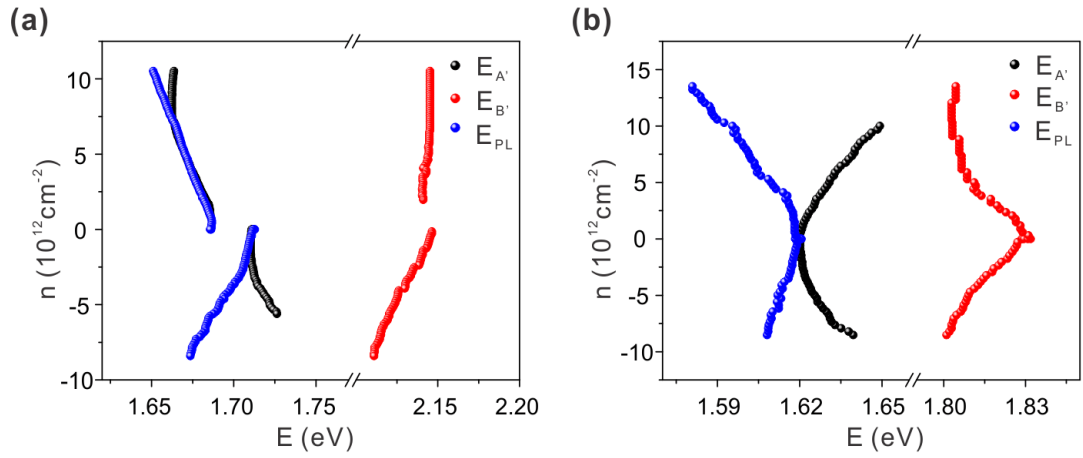

**Figure 4.** Relationship between the doping density and the peak energy of the A' and B' resonance features determined from the reflectance measurements and of the PL peak energy of the A' feature for monolayer $WSe_2$ (a) and $MoSe_2$ (b).